# EVIDENCE FOR SPIN-TRIPLET SUPERCONDUCTING CORRELATIONS IN METAL-OXIDE HETEROSTRUCTURES WITH NON-COLLINEAR MAGNETIZATION

Revised 6/20/2014 15:40:00

Yu.N. Khaydukov[1,4,6], G.A. Ovsyannikov[2,3], A.E. Sheyerman[2], K.Y. Constantinian[2], L. Mustafa[1], T. Keller[1,4], M.A. Uribe-Laverde[5], Yu.V. Kislinskii[2], A.V. Shadrin[2,3], A. Kalabukhov[3], B. Keimer[1], D. Winkler[3]

[1]Max-Planck Institute for Solid State Research, 70569, Stuttgart, Germany
[2]Kotel'nikov Institute of Radio Engineering and Electronics of Russian Academy of Sciences, 125009, Moscow, Russia.
[3]Chalmers University of Technology, Department of Microtechnology and Nanoscience, Gothenburg, S-41296, Sweden.
[4]Max Planck Society Outstation at FRM-II, D-85747 Garching, Germany
[5]University of Fribourg, Department of Physics, and Fribourg Centre for Nanomaterials, CH-1700 Fribourg, Switzerland
[6]Skobeltsyn Institute of Nuclear Physics, Moscow State University, Moscow, Russia

Heterostructures composed of ferromagnetic $La_{0.7}Sr_{0.3}MnO_3$, ferromagnetic $SrRuO_3$, and superconducting $YBa_2Cu_3O_x$ were studied experimentally. Structures of composition $Au/La_{0.7}Sr_{0.3}MnO_3/SrRuO_3/YBa_2Cu_3O_x$ were prepared by pulsed laser deposition, and their high quality was confirmed by X-ray diffraction and reflectometry. A non-collinear magnetic state of the heterostructures was revealed by means of SQUID magnetometry and polarized neutron reflectometry. We have further observed superconducting currents in mesa-structures fabricated by deposition of a second superconducting Nb layer on top of the heterostructure, followed by patterning with photolithography and ion-beam etching. Josephson effects observed in these mesa-structures can be explained by the penetration of a triplet component of the superconducting order parameter into the magnetic layers.

## I. INTRODUCTION

Recent research has shown that long-range triplet superconducting correlations can occur in a nonuniformly magnetized ferromagnet (F) in contact with a singlet superconductor (S) [1,2]. In S/F/S structures with uniform magnetization, the projection of the spin of a superconducting pair on the direction of the magnetization is conserved, and only singlet and triplet superconducting correlations with zero spin projection can appear [2,3]. Such superconducting correlations penetrate into the F layer over a length $\xi_F$, which is determined by the magnetic exchange energy $E_{ex}$ and typically amounts to several nm. In the dirty limit, $\xi_F = \sqrt{\hbar D/E_{ex}}$, where $D = v_F l/3$ is the diffusion coefficient, $v_F$ is the Fermi velocity, and $l$ is the mean free path. For magnets with nonuniform magnetization, triplet superconducting correlations can be generated in at the S/F interface, and their penetration length inside the magnet is predicted to be determined by the temperature $T$ as $\xi_N = \sqrt{\hbar D/k_B T}$, analogous to superconducting contacts with nonmagnetic metallic layers. Since the condition $k_B T \ll E_{ex}$ is usually satisfied in experiments, the appearance of long-range triplet superconducting correlations in a ferromagnet leads to an anomalously long-range proximity effect, manifested by superconducting currents in S/F/S Josephson junctions with thick ferromagnetic barriers [1-4].

The first experimental indications of the anomalously long-range proximity effect explained by the generation of long-range triplet superconducting correlations were obtained when studying an Andreev interferometer with a Ho film bridge with spiral magnetization [5] and critical current in S/F/S structures with $CrO_2$ layers [6,7]. These findings were confirmed in subsequent studies of single-crystalline Co nanowires [8], S/F/S structures with Heusler alloys [9], and a magnet with spiral magnetization [10]. Long-range spin-triplet superconducting currents were also observed in Josephson junctions containing composite magnetic layers that generate noncollinear magnetization between a central Co/Ru/Co synthetic antiferromagnet and two outer thin F layers [11,12]. A change in the superconducting critical temperature of a S/F/F' structure with a bilayer composed of two ferromagnets with non-collinear magnetization has also been reported [13,14].

All the above-mentioned studies were conducted on samples with elemental-metal or simple oxide layers, such as $CrO_2$. At the same time, using complex oxide perovskites as S and F layers brings some advantages. First, these compounds share similar crystal structures, which enable the preparation of epitaxial heterostructures with high quality of the layers and interfaces. Second, parameters such as the magnetic exchange energy can be tuned by changing the doping level of the complex oxides [13]. Third, critical temperatures of the copper-oxide superconductors are more than an order of magnitude larger than those of elemental superconductors, and thus of greater potential interest for spintronics applications. Several groups have contributed to the search for long-range triplet superconducting correlations in S/F/S structures with a manganite ferromagnetic layer with 100% spin polarization, where singlet superconducting correlations cannot appear. However, these studies have given contradictory results. On the one hand, the authors of Refs. [14,15] reported evidence of long-range triplet superconducting correlations from Andreev reflections in structures with a $La_{0.7}Ca_{0.3}MnO_3$ ferromagnetic layer. Other experiments on similar structures, however, did not reveal superconducting currents (beyond those transmitted through pinholes) [16,17].

In most of the studies mentioned above, the presence of non-collinear magnetization in the F layer was inferred from total magnetic moment measurements. For the direct observation of the non-collinear moments in composite F layers, polarized neutron reflectometry (PNR) can be used. In PNR, the intensities of specularly reflected neutron beams with different polarization $R^{\mu\eta}(Q)$ are measured. Here $Q = 4\pi\sin(\theta)/\lambda$ is the momentum transfer, and $\theta$ and $\lambda$ are the grazing-incidence angle and the neutron wavelength, respectively. An external magnetic field $H$ is typically applied perpendicular to the scattering plane and parallel to the heterostructure surface. The indices $\mu$ and $\eta$ take values "+" or "-" and correspond to the projection of the neutron spin parallel to $H$ before and after the scattering process, respectively. The non-spin-flip (NSF) reflectivities $R^{++}$ and $R^{--}$ depend on the depth profiles of the scattering length densities (SLD) $\rho^+(z)$ and $\rho^-(z)$ correspondingly. The latter depend on the nuclear SLD $\rho_0$ and in-plane component of the magnetic induction $B_\parallel$ parallel to $H$ as $\rho^\pm(z) = \rho_0(z) \pm cB_\parallel(z)$. Here the scaling factor is $c = 0.23\times10^{-4}$ kGs$^{-1}$nm$^{-2}$. The presence of an in-plane component of the magnetic induction $B_\perp(z)$ that is not aligned with $H$ leads to spin-flip (SF) scattering. The SF reflectivities $R^{+-}$ and $R^{-+}$ depend only on the magnetic potential $\rho^{SF}(z) = B_\perp(z)$. PNR thus allows the determination of depth profiles of the vector magnetization, and the experimental definition of the level of magnetic non-collinearity, which is an important parameter in the theory of triplet superconductivity [18-23].

PNR has been already used in study of magnetic state of oxide [24-26] and elemental metallic S/F structures [27,28]. In Ref. [12] PNR was used for the determination of the room temperature vector magnetization profile in a S/F/S Josephson system composed of an antiferromagnetically (AF) coupled Co/Ru/Co magnetic subsystem. A strong increase of the SF scattering, caused by the spin-flop transition of the AF coupled Co layers, was observed after subjecting the samples to a large in-plane magnetic field. This increase of the SF scattering was correlated with a 20-times enhancement of the superconducting critical current. The relatively low intensity of the SF scattering ($10^{-4}$ - $10^{-3}$ of the intensity of the direct beam) in conventional PNR experiments does not allow measurements of the temperature and field evolution of the vector magnetic profile within a reasonable time. In order to significantly increase the intensity of the SF scattering, we used waveguide enhancement of neutron standing waves [29,30] by forming a well-like structure of depth profile of SLD. Such a shape of the SLD depth profile allows trapping of neutrons inside the structure at certain values of the momentum transfer $Q_{wg}$, which leads to a $10^1$ - $10^2$ enhancement of the intensity of SF scattering. This allows a more detailed study of the non-collinear magnetic state.

The superconducting current through a composite layer with non-collinear magnetization was analyzed theoretically in Refs. [18-23]. In a trilayer geometry, the first F layer helps to convert singlet Cooper pairs into triplet pairs with non-vanishing projection onto the channels with parallel electron spins along the (tilted) magnetization of the central F layer, which may thus propagate coherently over long distances. The last F layer converts the triplet component back into the singlet state. Indeed, recent experiments observed a strong enhancement of the superconducting current through a ferromagnetic multilayer when the layers were ordered non-collinearly [10,11]. Recently it has been shown theoretically that a long-range triplet proximity effect may also develop in superconducting structures with a ballistic bilayer ferromagnet with non-collinear magnetization [19,20]. In this case, a second-harmonic Josephson relation is generated by the long-range propagation of triplet correlations which may then recombine into singlet Cooper pairs. The diffusive limit of superconducting structures with two ferromagnetic layers with non-collinear magnetization was considered in Refs. [20-23].

In this work, we experimentally study epitaxial hybrid heterostructures of composition Au/M/S, where Au is a thin film of gold, S is the cuprate superconductor $YBa_2Cu_3O_{6+x}$ (YBCO), and M consists of two thin layers of the ferromagnets $La_{0.7}Sr_{0.3}MnO_3$ (LSMO) and $SrRuO_3$ (SRO). The layer of gold on top is needed to prevent degradation of the system and also helps to create the neutron waveguide structure. The magnetization vector of the LSMO epitaxial film lies in the plane of the substrate [31], whereas the magnetization vector of the SRO film was directed at an angle of about 23° from the normal to the plane of the substrate [32]. An in-plane magnetic field of about 1T is needed to turn the vector of magnetization of the SRO layer collinear to the magnetization of the LSMO layer [31]. Recent PNR experiments on LSMO/SRO bilayers have revealed non-collinear magnetic order resulting from a competition between the magneto-crystalline anisotropy and the antiferromagnetic exchange coupling across the interface [33-35]. The previous transport study of LSMO/SRO bilayers sandwiched between two superconducting Nb and $YBa_2Cu_3O_{7-\delta}$ layers indicated the presence of the Josephson current in systems with a total thickness of the LSMO/SRO bilayer more than $\xi_F$ [36]. In the present study PNR and SQUID magnetometry are used for the quantitative description of the non-collinear magnetic state of the LSMO/SRO bilayer in combination with the transport measurements for the detecting the spin-triplet correlations generated by this noncollinear state. Section 2 of this article describes the fabrication technique and the experimental methods used in our study. In Section 3, we discuss X-ray data on the heterostructure. We then present the results of a characterization of the depth profile of the vector magnetization using SQUID magnetometry (Section 4) and PNR (Section 5). Section 6 contains results of transport measurements on mesa-structures. Section 7 provides conclusions of the work.

## II. SAMPLE FABRICATION AND EXPERIMENTAL TECHNIQUES

Heterostructures with composition $La_{0.7}Sr_{0.3}MnO_3/SrRuO_3/YBa_2Cu_3O_x$ (LSMO/SRO/YBCO) were fabricated on either (110) $NdGaO_3$ (NGO), (001) $LaAlO_3$ (LAO) or (001) $(LaAlO_3)_{0.3}(Sr_2AlTaO_6)_{0.7}$ (LSAT) substrates by pulsed laser ablation at temperature 700-800 °C and oxygen pressure 0.3-0.6 mbar. The heterostructures were covered by Au films in-situ after cooling to 100 °C. The thicknesses of the layers were as follows: 90-100 nm YBCO, 5-20 nm SRO, 5-30 nm LSMO, and 20 nm Au. Square mesa-structures with in-plane size $L$=10-50 μm were fabricated on (110) $NdGaO_3$ substrates [36, 37]. The lower electrode was an epitaxial film of YBCO, and the upper superconducting electrode was a Nb/Au bilayer. The layer M comprised two ferromagnets: SRO and LSMO (see inset in Fig. 1b). Peaks of all three materials of the heterostructures, YBCO, LSMO, and SRO, were observed in X-ray data of the LSMO/SRO/YBCO heterostructure. Magnetic measurements were conducted using a SQUID magnetometer in the temperature range from 10 to 300 K. The structural properties of the samples were determined by X-ray low-angle (reflectivity) and high-angle diffraction patterns on a Rigaku diffractometer with rotating anode. The magnetic field dependent DC resistance was measured on the mesa-structures patterned using photolithography and plasma chemical and ion etching [17,36,37]. Josephson current in mesa-structures was measured using four-point measurement scheme and magnetic field shielding by amorphous μ-metal foil in microwave screened environment. Microwave characteristics were determined from investigations of Shapiro steps, which arise in the *I-V* curves of mesa-structures irradiated by electromagnetic waves of frequency *fe*.

The PNR experiment was conducted on the angle-dispersive reflectometer NREX at the research reactor FRM-II in Garching, Germany. A polarized neutron beam with wavelength $4.26 \pm 0.06$ Å and polarization 99.99% falls on the sample under grazing incidence angles $\theta_1 = [0.15 - 1]°$. The divergence of the beam was set to $\Delta\theta_1 = 0.025°$ by two slits before the sample. The polarization of the reflected beam was analyzed by a polarization analyzer with efficiency 98%.

## III. X-RAY ANALYSIS OF THE HETEROSTRUCTURE

The crystal structure of Au/LSMO/SRO/YBCO

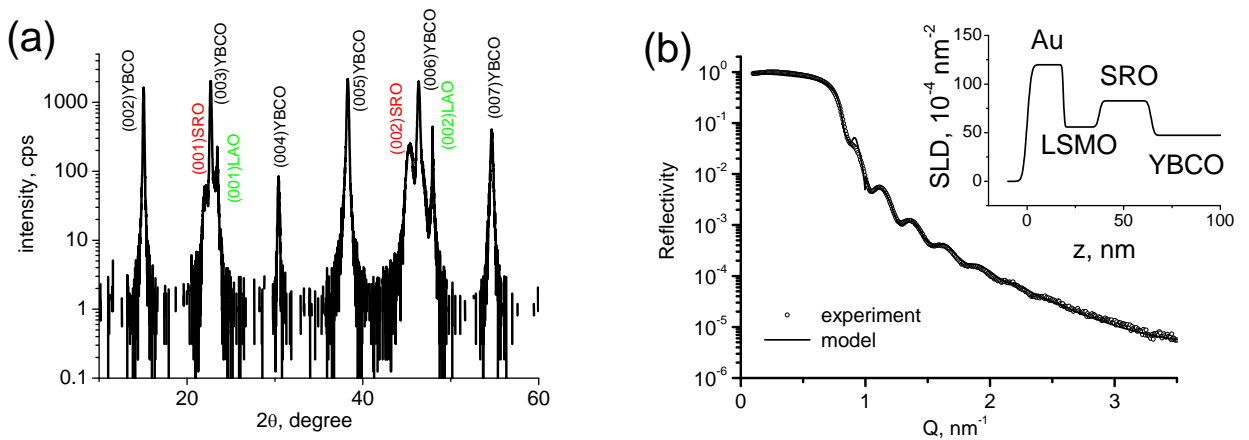

FIG. 1. (Color online) (a) X-ray 2θ-ω scan for the heterostructure Au/LSMO/SRO/YBCO deposited on a (001) $LaAlO_3$ substrate. (b) X-ray experimental (dots) and model (solid line) reflectivities for the same sample. Inset: X-ray SLD profile corresponding to the model curve.

heterostructures was investigated by X-ray diffraction. A $2\theta$-$\omega$ diffraction scan for the heterostructure LSMO/SRO/YBCO deposited on a (001) LAO substrate is presented in Fig. 1a. Since the substrate was slightly miscut, we observed sharp peaks with relatively low intensity at angles $2\theta = 23.4°$ and $2\theta = 47.9°$ corresponding to the Bragg reflections from the (00n) planes (n=1,2) of the pseudocubic (001) LAO substrate with lattice parameter $a_{LAO} = 0.389$ nm. Intense peaks from the (00n) planes (n = 2-7) of YBCO give interplanar distances $a_{YBCO} = 1.175$ nm. This value is bigger than lattice c-parameter of fully oxygenated YBCO deposited directly on the NGO substrate (see table 1), and indicates an oxygen stoichiometry $x \sim 0.6$. In addition, peaks from (001) SRO layer are shifted to the higher angles indicating a decrease in the interplanar distance of $a_{SRO} = 0.399$ nm in comparison with that of the SRO film deposited directly on the substrate. Peaks from LSMO are not discernible, since the position of the (002) LSMO peak coincides with the (006) peak of YBCO and the thicknesses of the LSMO layers are one order smaller than those of YBCO. However, additional experiments with LSMO/SRO heterostructures demonstrated that LSMO films deposited on top of SRO on NGO substrates are not strained, with $a_{LSMO} = 0.390$ nm [31]. Measurements of rocking curve at position of the (005) YBCO peak show a full-width at half maximum (FWHM) of 0.05° for the best YBCO films deposited on (110) NGO substrates (see table 1).

Table 1 Crystallographic parameters of the films and heterostructures

| Structure | Interplanar distance, nm (rocking curve FWHM, degree) | | | | |
|---|---|---|---|---|---|
| | NGO | LAO | YBCO | SRO | LSMO |
| YBCO/NGO | 0.3864 | - | 1.170 (0.3) | - | - |
| SRO/NGO | 0.3862 | - | - | 0.394 | - |
| LSMO/NGO | 0.3864 | - | - | - | 0.390 (0.04) |
| LSMO/SRO/ YBCO/LAO | - | 0.379 | 1.175 | 0.399 (0.25) | 0.391 (0.3) |
| LSMO/SRO/ YBCO/NGO | 0.3864 | - | 1.170 (0.1) | 0.399 (0.3) | 0.390 |

To check the quality of layers and interfaces, the X-ray reflectivity has been measured (Fig. 1b). The reflectivity curves are characterized by the presence of a reflection plateau at low angles, and Kiessig oscillations caused by the interference of reflections from different interfaces inside the structure. Fit of the experimental reflectivity to model curve allowed us to obtain information about the thicknesses of the layers and the root-mean-square (r.m.s.) roughness $\sigma$. The SLD depth-profile of the heterostructure that corresponds to the best agreement between experiment and model is depicted in the inset in Figure 1b. As follows from the fit, the LSMO/SRO and SRO/YBCO interfaces are sharp with the transition region less than 1 nm. The surface of the sample, in contrast, is rather rough, with $\sigma = 1.6$ nm. However, this does not influence the magnetic and superconducting properties of the system.

## IV. MAGNETIZATION

Magnetic measurements (Fig. 2) were conducted using a SQUID magnetometer in the temperature range from 10 K to 300 K in magnetic fields applied parallel (Fig. 2a and Fig.2b) and normal (Fig. 2c and Fig. 2d) to the surface. The measurements allowed the determination of the Curie temperatures of the LSMO ($T_M^{LSMO} \approx 350$ K) and of the SRO ($T_M^{SRO} \approx 130$ K) layers and superconducting transition temperature of YBCO layer ($T_C \approx 60$ K). The reduced $T_C$ value is in agreement with resistive measurements and with the c-axis lattice parameter of 1.175 nm measured by X-ray diffraction.

The temperature dependence of the in-plane magnetic moment $m_{//}(T)$ was measured by heating the sample from 10 K to 300 K after cooling in magnetic fields 30 Oe – 3000 Oe (Fig. 2a). For $T > T_M^{SRO}$ only the LSMO layer contributes to the magnetization of the sample. The experimental curve $m_{//}(T)$ follows approximately the predictions of mean field theory [38].

At $T_C < T < T_M^{SRO}$ the magnetic moment deviates from the theoretical curve due to the contribution of the in-plane component of the SRO magnetic moment. We have observed that depending on the magnetic field applied during cooling, the contribution to $m_{//}(T)$ can be either antiferromagnetic (for $H < 500$ Oe) or ferromagnetic (for $H > 500$ Oe), in agreement with Ref. [35].

Magnetic hysteresis loops describing the LSMO layer magnetization reversal were obtained in the temperature range 10-300 K with the magnetic field swept within 10 kOe. The curves for two temperatures are presented in Fig. 2(b). Measurements of the hysteresis loop conducted above $T_C$ provide direct confirmation of the ferromagnetic properties of the M layer. Increases of both the coercivity and the saturation magnetization upon cooling below $T_M^{SRO}$

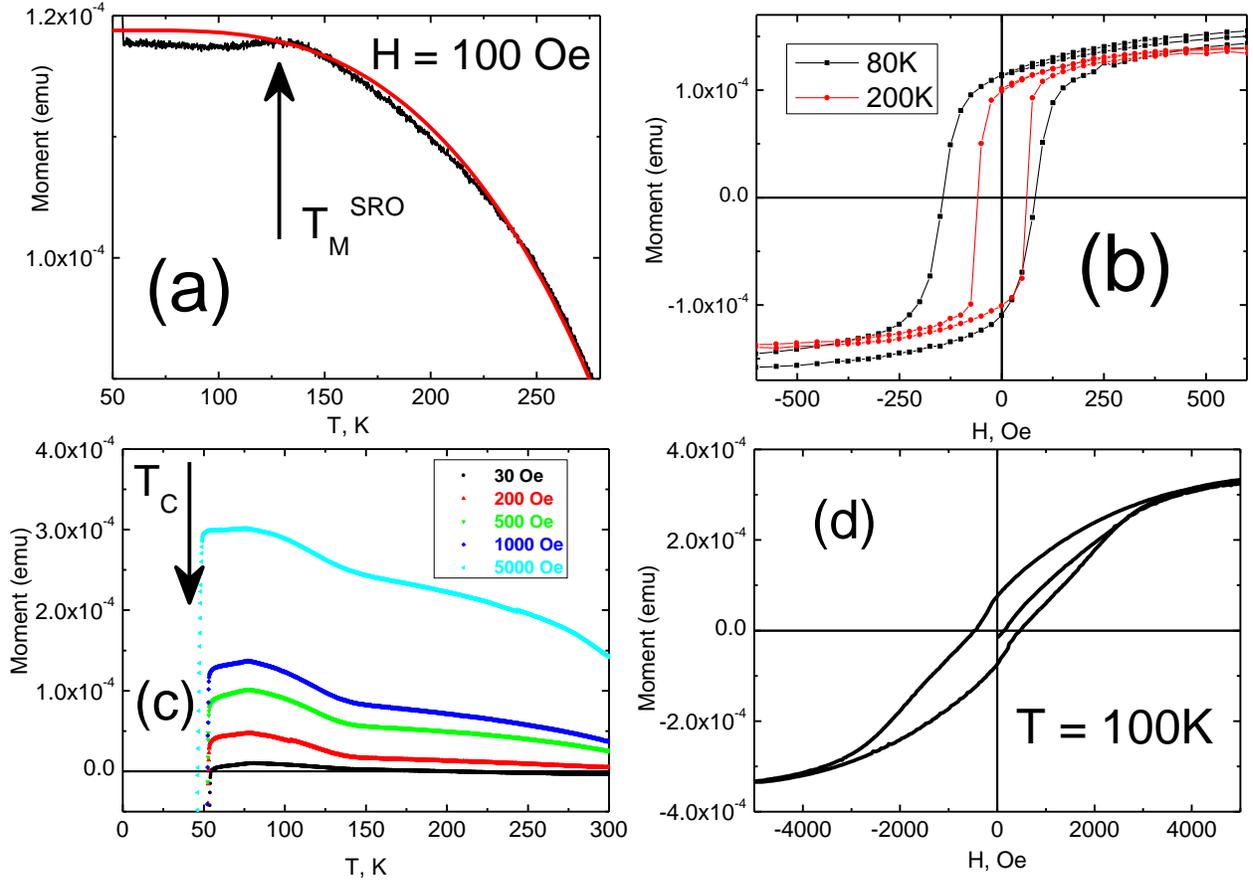

FIG. 2. (Color online) (a) Temperature dependence of the magnetic moment along magnetic field of Au/LSMO/SRO/YBCO/LAO heterostructure at magnetic field $H = 100$ Oe applied parallel to the sample surface. Solid line shows the theoretical dependence for LSMO layer within the mean-field approximation. (b) The in-plane hysteresis loop for the two temperatures. (c) Family of the $m_\perp(T)$ temperature dependencies measured at different magnetic fields applied normal to the sample surface. (d) The hysteresis loop measured at $T = 100$K for magnetic field applied normal to the sample surface

can be attributed to the transition of SRO to the ferromagnetic state.

To check the presence of an out-of-plane magnetic moment $m_\perp(T)$ of the SRO layer, we have conducted measurements with magnetic fields applied normal to the sample surface. The temperature dependence of out-of-plane magnetic moment $m_\perp(T)$ measured at different fields is presented in Fig. 2c. A significant increase of $m_\perp(T)$ is observed below $T_M^{SRO}$. However, Fig. 2c clearly shows a significant out-of-plane magnetic moment above $T_M^{SRO}$, which originates from the LSMO layer. To calculate the magnetic moment of the SRO layer we have subtracted the magnetic moment at $T < T_M^{SRO}$ from the moment slightly above $T_M^{SRO}$. The resulting moment for $H = 5$kOe is $6\times10^{-5}$ emu, which is somewhat smaller than the remanence moment $m_\perp$ measured at $T = 100$K (Fig. 2d). Based on the thickness of the SRO layer inferred from the X-ray reflectivity, we can estimate the out-of-plane magnetic moment of the SRO layer as 1.1-1.5 $\mu_B$/Ru.

## V. NEUTRON SCATTERING

Before the measurements the sample was cooled down to $T = 80$K in $H = 5$ kOe to align magnetic domains in the direction parallel to the external field. After this magnetic field was decreased to H = 30 Oe and reflectivity curves were measured. The next time, after cooling to $T = 80$K in $H = 5$ kOe, field was released to zero and the sample was cooled down to $T = 10$K. Reflectivity curves were measured then at constant temperature and different magnetic fields in the range from 30 Oe up to 5 kOe.

Spin-polarized reflectivity curves taken at $T = 80$ K are shown in Fig. 3a. The NSF curves $R^{++}(Q)$ and $R^{--}(Q)$ are characterized, similar to the X-ray data, by total reflection from the substrate with critical wave vector transfer $Q_{crit} = 0.15$ nm$^{-1}$ and by Kiessig fringes. The difference between $R^{++}$ and $R^{--}$ indicates the presence of a collinear component of the magnetization. The SF scattering, in turn, shows that an in-plane non-collinear component of the magnetization exists. The sharp peaks in the SF channels $R^{+-}$ and $R^{-+}$ at position $Q_{wg} \approx Q_{crit}$ with an intensity of about 10% of the intensity of the direct beam originate from the waveguide-like structure formed by capping the system with the layer of gold. The parameters of this peak (width, height and area) are very sensitive to the magnetic state of the system [29,30]. In particular, the magnetic field dependence of the peak area is shown in Fig. 3b.

To quantitatively describe the magnetic state of the system at a given temperature and magnetic field, we have fitted the experimental reflectivities $R_{exp}(Q)$ to model curves $R_{th}(Q)$ calculated using a supermatrix approach [39-41]. In the model, every layer was parameterized by the thickness $d$, nuclear SLD $\rho$, the root-mean-square (r.m.s.) roughness of the upper interface $\sigma$, the absolute value of the in-plane magnetization in the layer $M$, and the angle $\alpha$ between the $M$ and $H$. The fit of the model curves to the experimental data was made by minimizing the function $\chi^2 = 1/N \Sigma [(R_{exp} - R_{th})/\delta R_{exp}]^2$. Here $N$ is the total number of experimental points, and $\delta R_{exp}$ is the statistical error of $R_{exp}$. The summation runs over all experimental points. To find the global minimum of $\chi^2$ the "simulated annealing procedure" [42,43] was used. The thicknesses and r.m.s. roughness were taken from the fit of the X-ray reflectivity and kept fixed during the fit of the neutron data.

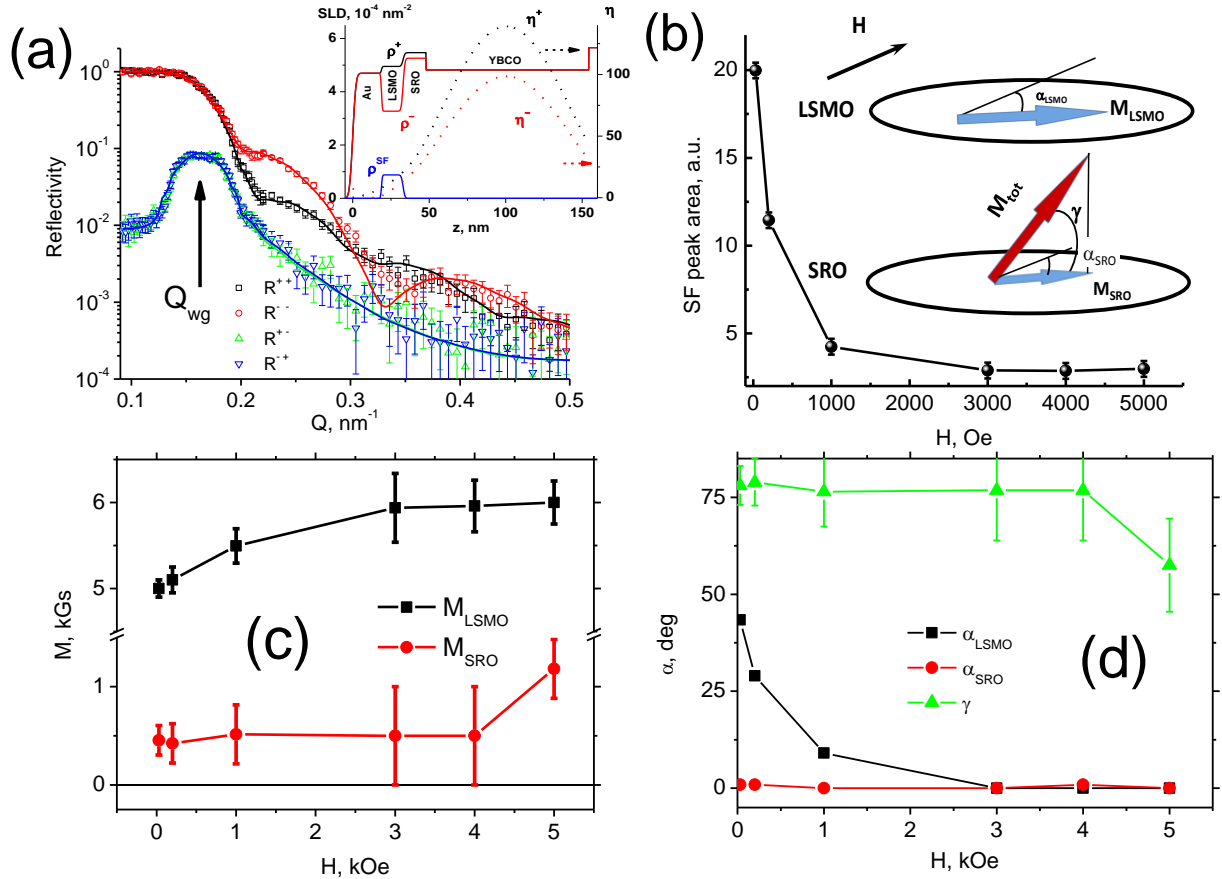

FIG. 3. (Color online) (a) Experimental (dots) reflectivity curves measured at $T = 80$K and $H = 30$ Oe. Model reflectivity curves are shown by solid lines. The vertical arrow shows the center of the waveguide peak. Inset: The SLD depth profiles correspondent to the model reflectivities (solid lines). Dashed lines show the density of spin up $\eta^+$ and spin down $\eta^-$ neutrons at the waveguide mode calculated for the correspondent SLD profiles. Note that $\eta^-$ is multiplied by factor of 4. (b) Integral of the waveguide spin-flip peak as a function of magnetic field measured at $T = 10$K. Inset: Sketch of the vector magnetic profile of the LSMO/SRO magnetic sub-system. Vector of magnetization of the LSMO layer lies in plane and makes angle $\alpha_{LSMO}$ with the external field. Vector of magnetization of the SRO layer is inclined on angle $\gamma$ to the sample plane. In-plane component $M_{SRO}$ is tilted on angle $\alpha_{SRO}$ to the direction of the external field. (d) The field dependence of the $M_{LSMO}$ (black) and $M_{SRO}$ (red). (e) The field dependence of the $\alpha_{LSMO}$ (black), $\alpha_{SRO}$ (red) and $\gamma$ (green).

The PNR curves at $T = 80K$ and $H = 30Oe$ were fitted by varying the SLDs of all layers, and the magnetic parameters of the LSMO and SRO layer. First, we tried a model where only the magnetic parameters of the LSMO layer $M_{LSMO}$ and $\alpha_{LSMO}$ were varied (see Fig. 3b). The smallest $\chi^2 = 3.44$ was obtained for $M_{LSMO} = 5 \pm 0.1$ kGs (corresponds to 2.5 $\mu_B$/Mn) and $\alpha_{LSMO} = 41.5°\pm0.3°$. The error of every parameter here is calculated as a 1% increase in the optimal $\chi^2$ and define the sensitivity of the fit to the given parameter. Knowing $d_{LSMO}$, $M_{LSMO}$ and $\alpha_{LSMO}$ and the area of the sample, $S = 25mm^2$, we can calculate the projection of total moment of the sample on external field in this model as $m \equiv S \times M_{LSMO} \times d_{LSMO} \times \cos(\alpha_{LSMO}) = 1.07 \times 10^{-4}$ emu. This value is almost 10% smaller than the one obtained by SQUID magnetometry. In the second model we also varied the magnetic parameters of the SRO layer, $M_{SRO}$ and $\alpha_{SRO}$. The best fit with $\chi^2 = 3.35$ is obtained for $M_{LSMO} = 5.0 \pm 0.1$kGs, $\alpha_{LSMO} = 43.3° \pm 0.3°$, $M_{SRO} = 0.5 \pm 0.1$ kGs and $\alpha_{SRO} = 1° \pm 2°$. The total magnetic moment in this case $m = 1.17 \times 10^{-4}$ emu agrees well with the SQUID data.

Having the $\rho^\pm(z)$ SLD profiles we can calculate the depth profiles of spin up $\eta^+(z)$ and spin down $\eta^-(z)$ neutron densities at the waveguide mode (inset in Fig. 3a). As it can be seen, the values $\eta^+(z)$ and $\eta^-(z)$ are 150 and 30 times enhanced in the middle of YBCO with respect to the density of the incoming neutron beam. The enhancement in the vicinity of the magnetic layers is of the order of 20-30. This enhancement allowed us to significantly increase the sensitivity of PNR in the determination of the in-plane non-collinear moment. For comparison, the sensitivity of the PNR curves at $Q > Q_{crit}$ to the determination of $\alpha_{LSMO}$ is only 2°, compared to 0.3° in the waveguide regime. We note that in Ref. [12] sensitivity of the PNR curves to the angles was 10° - 20°.

To fit the PNR curves measured at higher fields, we have only varied $M_{LSMO}$, $M_{SRO}$, $\alpha_{LSMO}$ and $\alpha_{SRO}$. The field dependence of the obtained parameters is shown in Fig. 3c and Fig. 3d. As it follows from Fig. 3c and Fig. 3d the decrease of the SF scattering depicted in Fig. 3b is mainly caused by the rotation of the magnetization vector of the LSMO layer from the direction of the easy axis (around 45° to the sample edge) towards external field. The saturation magnetization of the LSMO layer is 6.0 kGs (3.2$\mu_B$/Mn) is in good agreement with 2.9 $\mu_B$/Mn reported in Ref. [35] and 3.2 $\mu_B$/Mn reported in Ref. [34]. The calculation shows that the in-plane component of the SRO magnetization changes from 0.5kGs (0.3$\mu_B$/Ru) in $H = 30Oe$ up to 1.2kGs (0.7$\mu_B$/Ru) at $H = 5kOe$. Knowing the total magnetic moment of Ru (1.3 ± 0.2 $\mu_B$/Ru) from the SQUID data, we can calculate field dependence of the angle between magnetization vectors of LSMO and SRO $\gamma(H)$ (Fig. 3d). It can be seen that the non-collinear alignment of the LSMO and SRO magnetization vectors remains virtually unchanged in the range of applied magnetic fields $H = [0 \div 5]$ kOe that enables generation of a triplet condensate.

## VI. JOSEPHSON CURRENT IN MESA-STRUCTURES

To probe possible triplet superconducting correlations in the ferromagnetic layer, a mesa-structure with two superconducting electrodes was studied. The second electrode was a Nb film deposited on top of the Au/LSMO/SRO/YBCO structure [36]. Temperature dependence of resistance $R$ of the mesa-structure is shown in Fig. 4a. Two steps at $T_C^{YBCO}$ = 84.8K and $T_C^{Nb}$ = 8.7K corresponding to the superconducting transition of YBCO and Nb layers are clearly seen on the $R(T)$ dependence. A critical current $I_C$ with linear temperature dependence was observed below $T_C^{Nb}$ (Fig. 4b).

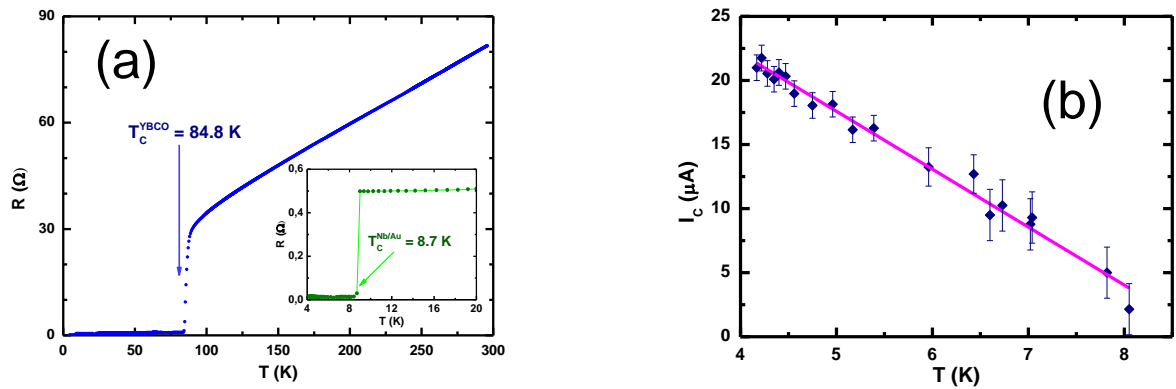

FIG. 4. (Color online) (a) Temperature dependence of resistance $R(T)$ of mesa-structure with $d_{SRO} = 10nm$ and $d_{LSMO} = 9nm$. Inset: $R(T)$ dependence in the vicinity of transition to superconducting state of Nb-Au electrode. (b) Temperature dependence of critical current for the same sample.

A superconducting current was observed in all mesa-structures with LSMO/SRO composite ferromagnetic bilayer with total thickness $d_M = d_{LSMO} + d_{SRO}$ up to 53 nm (Fig. 5). This is much larger than the coherence length of the ferromagnets $\xi_F \sim$ 5nm estimated for LSMO and SRO films [36]. Control measurements of the mesa-structure with only the LSMO [17] or the SRO layer [36] showed that the critical current is absent (except in cases of pinholes) if the SRO and LSMO films are thicker than several nm. The critical current density $j_C$ decreases by an order of magnitude, when increasing $d_M$ from 8.5 to 53 nm. The maximum $j_C$ = 90 A/cm$^2$ was observed for the sample with $d_{LSMO}$ = 6 nm and $d_{SRO}$ = 8.5 nm having surface area of 100 μm$^2$ Note that a non-monotonic $j_C(d_M)$ dependence with a maximum at $d_M \sim \xi_F$ was predicted in Ref. [22] for structures with long-range triplet superconducting correlations. Further increase in $d_M$ resulted in a decrease of $j_C$ as expected from theoretical calculations [21-23].

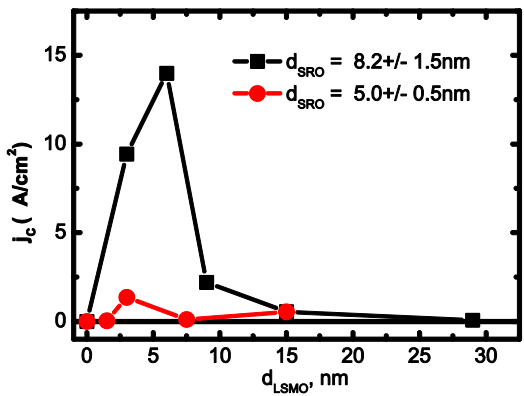

FIG. 5. (Color online) Dependence of the critical current density $j_C$ measured at $T$=4.2K on $d_{LSMO}$ at $d_{SRO}$ = 8.2nm (black dots) and $d_{SRO}$ = 5.0nm (red circles). For each point the data was obtained by averaging over several samples with different surface areas.

The measurements of the critical current $I_C$ as a function of relatively small magnetic field $H \sim$ 10 Oe was already reported in Ref. [36]. Subsequent measurements showed that critical current exists even in the fields of several kOe, where non-collinear alignment of magnetization in LSMO and SRO still exists (see previous sections). Note, it should be surprising for singlet superconducting current to exist at $H$ fields being up to 100 times stronger than the period of critical current oscillation which was of order of 10- 20 Oe.

The absence of pinholes in samples under the test was confirmed by structural, magnetic and microwave measurements. Indeed, presence of the pinholes in the investigated samples would lead to the deviation of structural and magnetic properties. However, as it was shown above, X-ray and neutron reflectometry and SQUID measurements show that structural and magnetic properties of the LSMO and SRO layers are close to the literature values.

Another check on the absence of the pinholes is the analysis of Shapiro steps on current-voltage (*I-V*) characteristic under microwave irradiation. It's important to note that autonomous *I-V* curves are less informative and a nice looking *I-V* curve measured at dc may belong to a mesa-structure with pinholes. As a rule of thumb, impact of pinholes on microwave dynamics of Josephson junction resulted in significant reduction of Shapiro steps heights from expected ones by the resistively shunted junction (RSJ) model [44].

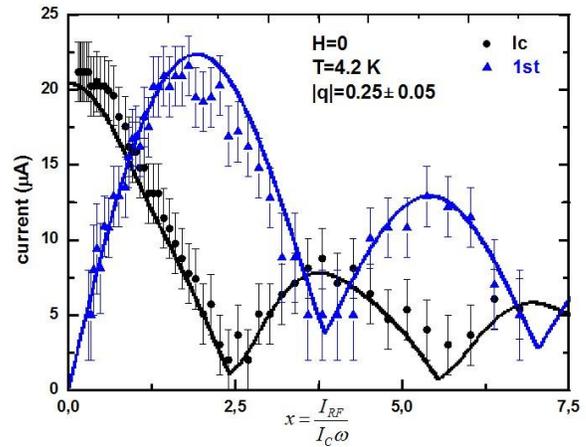

FIG. 6. (Color online) The critical current (circles) and amplitudes of the first (triangles) Shapiro steps as functions of the normalized amplitude of microwave signal for the sample with $d_{SRO}$ = 10nm and $d_{LSMO}$ = 9nm. A fit (lines) to an expression calculated within the modified RSJ model yields 25% as the fraction of the second harmonic in the current-phase relation.

Fig. 6 shows that the maximum of the first Shapiro step is $I_1$ =21 μA and, correspondingly, the ratio of $I_1/I_C$ is in good agreement with the RSJ model ruling out thus presence of pinholes. Note, the zero field cooled I-V curves presented in Fig. 6 do not differ much from the field cooled ones at $H$ = 52 Oe measured even at somewhat higher temperature. In the case of the singlet superconducting pairing expected amplitudes of critical current and, correspondingly, the height of the principal Shapiro step would be significantly suppressed.

## VII. CONCLUSION

We have directly probed non-collinear magnetism on the metal-oxide heterostructures by means of SQUID magnetometry and polarized neutron reflectometry. The dependence of the observed superconducting current in the mesa-structures Nb/Au/LSMO/SRO/YBCO on thicknesses of LSMO and SRO layers has been studied and compared with theoretical predictions. The Josephson effect observed in these structures is explained by the penetration of the long-range triplet component of the superconducting correlations into the magnetic layer. Further work is required to elucidate the magnetic structures at the interfaces and their influence on the propagation of supercurrents, as well as the possible role of d-wave pairing in the YBCO layers.

## ACKNOWLEDGEMENTS

We are grateful to I.V. Borisenko, A. Buzdin, T. Claeson, G, Cristiani, V.V. Demidov, M. Fogelstrom, T. Lofwander, G. Logvenov, A.S. Mel'nikov, A.M. Petrzhik, A.V. Zaitsev for assistance in the experiment and stimulating discussions. This work was supported by the Division of Physical Sciences, Russian Academy of Sciences; the Ministry of Education and Science of the Russian Federation; the Council of the President of the Russian Federation for Support of Leading Scientific Schools (Grant NSh-4871.2014.2); the Russian Foundation for Basic Research (project. 14-07-00258); and the Swedish Institute (Visby program) and Deutsche Forschungsgemeinschaft within the framework of the TRR80 project. This work is based upon experiments performed at the NREX instrument operated by Max-Planck Society at the Heinz Maier-Leibnitz Zentrum (MLZ), Garching, Germany. The neutron part of the project has been supported by the European Commission under the 7th Framework Programme through the "Research Infrastructures" action of the Capacities Programme, NMI3-II, Grant Agreement number 283883.